# Advanced Virus Monitoring and Analysis System

Fauzi Adi Rafrastara
Faculty of Information and Communication Technology
University of Technical Malaysia Melaka
Melaka, Malaysia
fauzi_adi@yahoo.co.id

Faizal M. A
Faculty of Information and Communication Technology
University of Technical Malaysia Melaka
Melaka, Malaysia
faizalabdollah@utem.edu.my

*Abstract* — **This research proposed an architecture and a system which able to monitor the virus behavior and classify them as a traditional or polymorphic virus. Preliminary research was conducted to get the current virus behavior and to find the certain parameters which usually used by virus to attack the computer target. Finally, "test bed environment" is used to test our system by releasing the virus in a real environment, and try to capture their behavior, and followed by generating the conclusion that the tested or monitored virus is classified as a traditional or polymorphic virus**.

## I. INTRODUCTION

Nowadays, we all live in the digital era, which most information moves from one place to another digitally. The information can be derived easily from everywhere and send it to whoever, only in minutes or even seconds. Unfortunately, wherever we are, including in this digital information era, threats always exist, perhaps in the different shapes. One of the popular threats which always peering us in this era, is Computer Virus.

The virus is a threat, because it can do bad things to whomever. It can make the computer becomes slow, broken, or even it can delete the data. The virus can run automatically and hide the process, so that users cannot see the processes and activities, which are done by virus. What can users see from the virus is what they have done.

## II. BACKGROUND

There is a kind of software that can be used to detect the existence of Virus inside the computer, called Anti Virus (AV). AV is widely used to detect and combat the virus. They will report to the user when they found the virus inside the PC. Unfortunately, they cannot list and report all behaviors or activities of the virus [1]. This limitation of AV has been covered by the certain tools, which mostly do not have a capability in virus detection system, called Virus Monitoring and Analysis Tool (VMAS). VMAS is specially used to monitor and analyze as well as capture all activities performed by virus [1]. VMAS also can generate the details report regarding the virus's behavior. This kind of report is important for those who want to learn more about virus activities. Furthermore, people can eliminate the viruses from their PC and recover the Operating System from viruses attack by reading the virus behavior analysis report [2]. There are several popular VMAS which mostly used to get the data of virus's behavior, such as CWSandbox, Capture, MBMAS, Joebox and ThreatExpert [2].

The aforementioned tools indeed are able to produce the behavior analysis report in details. Unfortunately, by using these tools, the type of malicious file, that have been tested, still cannot be recognized. Even though the analysis report can be derived, it is not easy to determine which virus file is classified as traditional or polymorphic only by reading this report [2][3].

However, either AV or VMAS cannot distinguish between traditional and polymorphic virus. They are only capable of detecting and reporting the virus behavior. Whereas, classifying the virus automatically, it will be a different task which has not been solved yet. So, in this research, a new architecture will be proposed as well as the system. This architecture and system are served to classify the virus automatically whether it is considered as a traditional or polymorphic virus.

## III. DATA COLLECTION

Data Collection is needed to conduct the preliminary research in which all the required data will be collected manually. Further, these data will be compared to the generated data in testing phase. Here 20 viruses will be examined and analyzed one by one. This step is important to classify whether these viruses are categorized as a traditional or polymorphic virus. Based on this manual experiment, two viruses were detected as a polymorphic virus, since they always obfuscated their signatures whenever they propagate [4] [5], as listed in Table I. The signature that was identified in our research here is MD5 checksum [6][7]. This kind of checksum is popular to be used by current antivirus to detect the existence of viruses based on their signatures [8][9][10].

Further, these data will be used to validate the final data which generated by the proposed system. The proposed system can be considered to be successful if it can produce the same result and conclusion with the data from this preliminary research.

TABLE I. LIST OF THE ANALYZED VIRUS

| No. | Virus Name | Detected by | Types |
|---|---|---|---|
| 1. | W32.Blaster.E.Worm (Lovesan) | Symantec | Traditional |
| 2. | W32.Downadup.B (Conficker) | Symantec | Traditional |
| 3. | W32.Higuy@mm | Symantec | Traditional |
| 4. | W32.HLLW.Benfgame.B (Fasong) | Symantec | Polymorphic |
| 5. | W32.HLLW.Lovgate.J@mm | Symantec | Traditional |
| 6. | W32.Imaut | Symantec | Traditional |
| 7. | W32.Klez.E@mm | Symantec | Polymorphic |
| 8. | W32.Kwbot.F.Worm | Symantec | Traditional |
| 9. | W32.Mumu.B.Worm | Symantec | Traditional |
| 10. | W32.Mytob.AV@mm | Symantec | Traditional |
| 11. | W32.SillyFDC (Brontok) | Symantec | Traditional |
| 12. | W32.SillyFDC (Xema) | Symantec | Traditional |
| 13. | W32.Sober.C@mm | Symantec | Traditional |
| 14. | W32.Swen.A@mm | Symantec | Traditional |
| 15. | W32.Valla.2048 (Xorala) | Symantec | Traditional |
| 16. | W32.Virut.CF | Symantec | Traditional |
| 17. | W32.Wullik@mm | Symantec | Traditional |
| 18. | W32/Rontokbro.gen@MM | McAfee | Traditional |
| 19. | W32/YahLover.worm.gen | McAfee | Traditional |
| 20. | Worm:Win32/Orbina!rts | Symantec | Traditional |

## IV. THE PROPOSED ARCHITECTURE AND SYSTEM

Since the main objective of this research is to propose an architecture and system which is able to classify between traditional and polymorphic virus, so this research focuses on the host side attack only.

In this research, two tools have been developed to classify between polymorphic and traditional virus, which are *Virus Behavior Monitoring Tools* (VBMT) and *Virus Behavior Analysis Tool* (VBAT). These tools are included in one system, called *Advanced Virus Monitoring and Analysis System* (AVMAS).

VBMT is served to monitor the activity of virus. They will execute the virus and then captured all activities which are performed by virus, during monitoring time. Usually current VMASes take maximal 4 minutes along for the monitoring time [1][11][12]. The VBMT will be installed into two PCs. Later, the same virus will be executed and monitored inside these PCs, to know whether or not the virus performs different things, especially in term of offspring's signature.

On the other hand, VBAT is used to analyze the results that generated by each VBMT. This analysis process is important to come up with the conclusion that the tested virus is classified as a polymorphic or traditional virus.

The proposed architecture here actually can be implemented in two environments, which are real environment and virtual environment. Real environment means, by providing at least two PCs to test the virus and installing VBMT into these PCs. One more PC is needed to be installed with a VBAT. Fig. 1 illustrates the architecture of AVMAS in real environment.

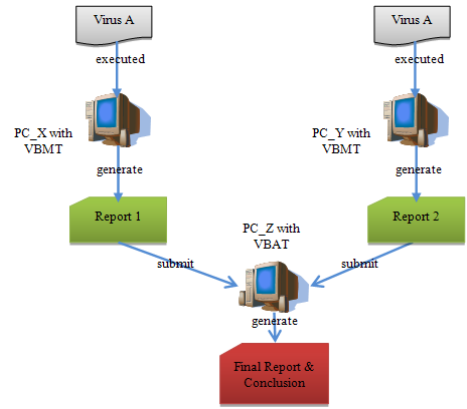

Figure 1. Architecture of AVMAS in real environment

The main concept of this architecture here is, a virus is tested in two PC with VBMT inside, by which VBMT will monitor and captured all activities which are performed by the tested virus. After monitoring time is finished, then each VBMT will generate a result that is reporting all activities captured, including new files generated and their checksums. Further, these two reports should be submitted to VBAT which installed inside the third PC. VBAT is tasked to analyze and compare between these two reports, and come up with the conclusion whether the tested virus is classified as a polymorphic or traditional virus. If VBAT found the fact that there is a difference between the first report with second report, especially in term of virus's activity or the signature of new files generated, so VBAT will conclude that the tested virus is classified as a polymorphic virus [1][4][5], otherwise it classified as a traditional virus [1][5].

This architecture actually can be simplified by using only one PC, but two virtual machines must be installed inside. The concept of the second architecture is almost similar to the first one. The difference is the location of VBMT which is installed inside these two virtual machines. Meanwhile, VBAT will be put inside the main or real PC. Fig. 2 shows the architecture of AVMAS in virtual environment.

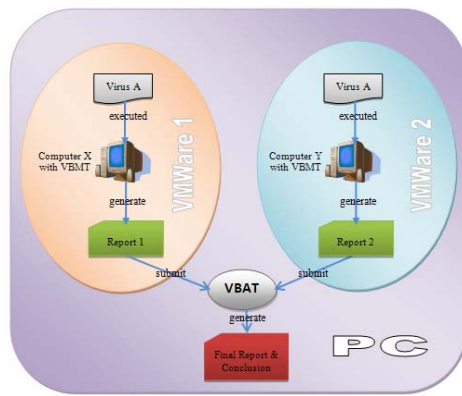

Figure 2. Architecture of AVMAS in virtual environment

## V. TESTING AND RESULT

After completing the development phase, testing process should be done to make sure that the proposed system can be used to deal with the problems. Fig. 3 shows the flowchart to test the AVMAS. Firstly, a virus is put into two PCs with VBMT inside. After that, the virus is executed and monitoring process is started. Monitoring process will be performed in 5 minutes along, because according to [1][11][12], usually current VMAS take maximal 4 minutes to monitor virus activity. During this monitoring process, all virus behaviors, especially which relating to host side effect will be captured. When the timeout limit have been reached, each VBMT will generate the report consisting all behavior captured and the required data to classify the tested virus whether it is considered as a traditional or polymorphic virus.

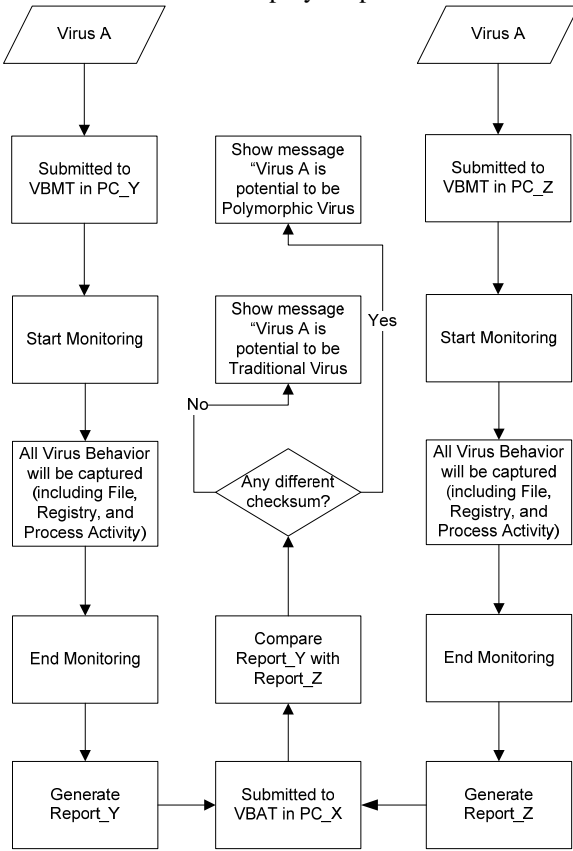

Figure 3. Flowchart to test the AVMAS

The next step is by submitting each report into VBAT which is installed in the third PC. This VBAT will compare the first report to the second report, especially in term of checksum generated. Once it finds the differences, so it means that, the tested virus can generate the different signature of offspring in the different PC. This conclusion addresses to the further conclusion that, this virus can be considered as a polymorphic virus.

On the other side, when the VBAT finds the same content between these two reports, including the generated checksums, so straight away VBAT will come up with the conclusion that this virus is classified as a traditional virus.

TABLE II. RESULT COMPARISON BETWEEN DATA FROM PRELIMINARY RESEARCH AND AVMAS TESTING

| No. | Virus Name | Preliminary Research Result | AVMAS Testing Result |
|---|---|---|---|
| 1. | W32.Blaster.E.Worm (Lovesan) | Traditional | Traditional |
| 2. | W32.Downadup.B (Conficker) | Traditional | Traditional |
| 3. | W32.Higuy@mm | Traditional | Traditional |
| 4. | W32.HLLW.Benfgame.B (Fasong) | Polymorphic | Polymorphic |
| 5. | W32.HLLW.Lovgate.J@mm | Traditional | Traditional |
| 6. | W32.Imaut | Traditional | Traditional |
| 7. | W32.Klez.E@mm | Polymorphic | Polymorphic |
| 8. | W32.Kwbot.F.Worm | Traditional | Traditional |
| 9. | W32.Mumu.B.Worm | Traditional | Traditional |
| 10. | W32.Mytob.AV@mm | Traditional | Traditional |
| 11. | W32.SillyFDC (Brontok) | Traditional | Traditional |
| 12. | W32.SillyFDC (Xema) | Traditional | Traditional |
| 13. | W32.Sober.C@mm | Traditional | Traditional |
| 14. | W32.Swen.A@mm | Traditional | Traditional |
| 15. | W32.Valla.2048 (Xorala) | Traditional | Traditional |
| 16. | W32.Virut.CF | Traditional | Traditional |
| 17. | W32.Wullik@mm | Traditional | Traditional |
| 18. | W32/Rontokbro.gen@MM | Traditional | Traditional |
| 19. | W32/YahLover.worm.gen | Traditional | Traditional |
| 20. | Worm:Win32/Orbina!rts | Traditional | Traditional |

Based on our test experiment, we found that this system can classify the tested virus correctly, with 100% similar to the data from preliminary research, as listed in Table II.

## VI. CONCLUSION AND FUTURE WORK

In the monitoring process, this research focused on the host side attack, in which consist of three parameters that should be monitored, such as file, registry, and process activity. Whereas, to analyze the result for virus classification, there are several parameters used in this research, which are file activity, especially executable file creation, by comparing their checksums which produced in one PC to the checksum from antoher PC.

In the data collection phase, the viruses' behavior and activity especially which related to the host side have been captured, either manually or by using the third-party tools, such as: Joebox and ThreatsExpert. This data is used to match the result obtained from AVMAS. The result of this test and validation process show that, the system called AVMAS is able to monitor and classify the tested virus with same conclusion than one generated manually.

For the future work, this research can be improved to be a system, which is not only able to classify between traditional and polymorphic virus, but also to classify metamorphic virus as well. Next, this research also can be developed further to produce a system that is able to monitor and analyze the activity of a virus, then produce the virus removal tool automatically. It will be very beneficial to common users who want to clean their computers, which have been infected by the virus, since antivirus focuses on the prevention side so far, rather than cure action.

ACKNOWLEDGMENT

This research was supported by University of Technical Malaysia Melaka (UTeM) and Fundamental Research Grant Scheme (FRGS).

AUTHORS PROFILE

**Fauzi Adi Rafrastara**. He was born in Semarang, Indonesia, 30 April 1988. He obtained the bachelor's degree from Dian Nuswantoro University, Indonesia, in 2009. He is currently a master student at University of Technical Malaysia Melaka. His research area include software engineering and information security.

**Faizal M. A**. He is currently a senior lecturer in University Technical Malaysia Melaka. He obtained the Ph.D degree in 2009 from University Technical Malaysia Melaka focusing on Intrusion Detection System. He research are IDS, Forensic, and Network Security.